\documentclass[vecphys]{svmult}

\usepackage{makeidx}         % allows index generation
\usepackage{graphicx}        % standard LaTeX graphics tool
\usepackage{multicol}        % used for the two-column index
\usepackage[bottom]{footmisc}% places footnotes at page bottom
\makeindex             % used for the subject index
\begin{document}

\title*{Disks around Young Binary Stars}
\author{L. Prato\inst{1}\and
A. J. Weinberger\inst{2}}
\institute{Lowell Observatory, 1400 West Mars Hill Road, Flagstaff,
AZ, 86001 \texttt{lprato@lowell.edu}
\and Department of Terrestrial Magnetism, Carnegie Institution of Washington, 5241 Broad Branch Road, NW, Washington, DC 20015; and NASA Astrobiology Institute
\texttt{weinberger@dtm.ciw.edu}}
\maketitle

Should this be an abstract?

\section{Introduction}
\label{sec:1}
% Always give a unique label
% and use \ref{<label>} for cross-references
% and \cite{<label>} for bibliographic references
% use \sectionmark{}
% to alter or adjust the section heading in the running head

It is an observational fact that among young stars in many nearby
star forming regions (SFRs) an excess binary population exists
(e.g., \cite{ghe93, lei93, sim95}).  This overabundance of doubles,
in comparison to field stars in the solar neighborhood \cite{duq91},
correlates at least 
with the property of stellar density \cite{pro94, pet98,
pat02, bec03}: the denser clusters contain a lower fraction of
bound multiple systems. The maximum separation of bound systems is
also related to the stellar density. \cite{sim97} used a two-point
correlation function to show that the transition between the
binary and large-scale clustering regimes, and hence in the
cutoff separation
for the likelihood of a bound pair, increases from 400~AU (Orion
Trapezium) to 5,000~AU (Ophiuchus) to 12,000~AU (Taurus), while the
average stellar surface density decreases.
Studies of large samples of binaries in very
different star forming regions are key to unravelling the nature
of binary formation mechanisms and the impact of environment
on multiplicity fraction, distribution, and evolution.

The frequency and separation of young binary populations
are perhaps most important when examined in light of
the impact of companion stars on the potential for
planet formation.  Even for star forming regions in which
the binary frequency is similar to that of the local
field population, roughly two thirds of all member
stars form in multiple systems. 
For a certain range of stellar separations,
the presence of a companion star will clearly impact 
the formation, structure, and evolution of circumstellar
disks, and, hence, of any potential planet formation.
An insoluble problem among main-sequence field stars
is the possibility of prior dynamical evolution of the system
(e.g., \cite{por05}).  The interactions between young
stars and their associated circumstellar and circumbinary
disks may set in motion this dynamical evolution.  However,
examining these young systems in particular tells us about
the initial \emph{potential} for planet formation.  Field
star observations tell us if this potential was realized.

For very small separation binaries, models indicate
that planet formation should be possible in a
circumbinary disk (e.g., \cite{qui06}).  Several
examples of close young binaries with circumbinary
disks are well known, such as DQ Tau e.g., \cite{mat97},
UZ~Tau~E \cite{pra02, mar05}, and HD~98800B \cite{koe00,
pra01}.  These pairs have separations of $\sim$30~R$_{\odot}$
to 1~AU \cite{bas97, pra02, bod05}.  The GG Tau and UY Aur
binaries, with stellar separations of tens of AU, are
surrounded by angularly large and therefore well-studied
circumbinary disks (e.g., \cite{mcc02, clo98}).
Recently, \emph{Spitzer} observations of main-sequence pairs
revealed debris disk material around 14 young stars with
separations of several solar radii to $\sim$5~AU \cite{tri07}.
However, in spite of these promising disk observations and
model predictions, no planet has yet been detected orbiting a
small separation, main-sequence binary (although a 2.4~M$_{Jup}$
minimum mass planet orbits the G6V star HD~202206 and its
a$=$0.83~AU substellar companion \cite{udr02}).  This dearth of
detections may simply reflect the difficulties inherent in radial
velocity (RV) searches for planets around binaries and the
fact that binaries are typically eliminated from RV samples
(e.g., \cite{egg04, kon05}).

Models also indicate a favorable outcome for planet
formation in the circumstellar disks of binaries \cite{qui07}.
Reservoirs for this process, the optically thick,
circumstellar disks
around component stars, are routinely observed in binary
systems with separations as small as $\sim$14 AU
(e.g., \cite{har03}).  More than 30 extrasolar
planets ($\sim$20\%) have been reported around one component in
binaries with separations of tens of AU up to thousands of AU
(\cite{egg04, rag06}) --- circumstellar planet
formation seems to be common in multiple systems.

Necessary truncation of the outer portions of
circumstellar disks in binaries with separations of a
few to several 10's of AU likely delineates a planet-free
zone.  Interestingly, this fiducial separation is similar to
that of the peak in the separation
distribution for binaries in most SFRs (e.g., \cite{pat02}).
This "planet-free" regime of binary separation is also notably
the least well-studied; components at such separations are too
distant to be observed as spectroscopic binaries (orbital
induced RV variations are on the order of star spot induced RV
variations \cite{Saar98}), yet too close to be easily angularly
resolved.  Few data sets that go beyond initial binary
identification exist, although there are some exceptions
such as \cite{har03}.  We therefore loosely
define the binary separation regime most interesting,
under-studied, and potentially
treacherous to the formation and longevity of circumstellar
disks, and therefore to the of formation of planets, as
spanning a few AU to 30~AU.  This definition is naturally
modulo eccentricity and mass ratio, properties which could
reinforce circumstellar disk destruction on short time scales.

In this paper, we will discuss the current state of observations
of disks in young multiple systems with an emphasis on
circumstellar structures.  Disks in solar analogue and low-mass
stellar systems will be primarily considered.
The topics covered in this review are
the evolution of inner disks in binaries (\S 2), the evolution
of outer disks in binaries and the determination of disk
masses as derived from submillimeter astronomy(\S 3), the
orientation of disks in binary systems (\S 4), and the
structure of debris disks in young binaries (\S 5).
We will present these topics through the lens of the
potential for planet formation in these systems.
In summary, \S 6 will present a discussion of future
experiments and observations required to move knowledge
in this field forward.

\section{Inner Disks}
\label{sec:2}

Hydrogen emission line
diagnostics (H$\alpha$ or Br$\gamma$) and near-infrared colors
are effective determinants of weak-lined (no inner disk) and classical
(optically thick inner disk) young stars (see \cite{mar98,
pra97}).  Substantial line emission and near-infrared
excesses attest to the presence of gas and warm dust located in
the inner $\sim$2~AU of a circumstellar disk around a
G -- M spectral type young star.  These disks are thought to
evolve quickly from optically thick to thin states; few systems have
been found in the intermediate ``transition'' state. The inner few
AU of a circumstellar disk is the likely site of terrestrial planet
formation and thus is particularly important.

Monin et al. (2007) classified a sample of young binaries
with separations of $\sim$15 -- 1500 AU on the basis of these
diagnostics.  In an extensive search of the young star binary
literature, only $\sim$60 systems were found for which both
component spectral types were known and for which angularly
resolved H$\alpha$, Br$\gamma$, K--L, or K--N (K$=$2.2$\mu$m,
L$=$3.4$\mu$m, and N$=$10$\mu$m) color data were
available.  These few dozen systems are drawn from a variety of
star forming regions, and thus do not represent a homogeneous
sample.  This dramatically underscores the unavoidable small
number statistics inherent in any analysis of this sample, and the
pressing need for a substantial observational effort in this area.

In spite of the small sample size,
\cite{mon07}'s analysis revealed intriguing results and
trends (Figure 1).  One surprising and relatively robust
outcome is that mixed pairs, in which the
components appear to be in different evolutionary stages, are not as
rare as once believed (e.g., \cite{pra97, har03}), comprising
approximately 40\% of the sample.  Less statistically notable are
the suggestions that these system are more common
among the larger separation pairs and that a slight majority
these systems are detected among the lower mass ratio pairs.
There is also a hint in the available data that the frequency
of mixed pairs may vary between star forming regions.
Unfortunately, because of the sparse data, these results are
all at the 2$\sigma$ level at best.

Angularly resolved high-resolution spectroscopy of close young
binaries yields insight
into either the alignment of stellar orbital axes, or
discrepant rotation rates.  This degeneracy can be resolved
with a time series observations designed to obtain component
stellar rotation periods.  If rotation axes are aligned,
discrepant rotation periods would suggest star-disk
locking in only one component.  What regulates such
discrepancies in double star systems that presumably form
and evolve together in the same relative environment?
Determination of the stellar properties favorable to long-lived
inner disks bears directly on the question of what kind of stars
are most likely to host planets.  Figure 1 shows a young binary
with component $v$sin$i$'s discrepant by a factor of 2$-$3.
Intriguingly, this $\sim$700~AU separation Ophiuchus binary, an
M3 and an M7, is a mixed system (e.g., \cite{pra03}).  The
rapidly rotating primary is not associated with dusty
circumstellar material, however, the low mass companion is
\cite{mcc06}, as we might expect from a disk-locking scenario.
Similar discrepancies have also been observed in a 30~AU
separation young binary in Taurus (results in preparation for
publication).

\begin{figure}
\centering
\includegraphics[height=9cm]{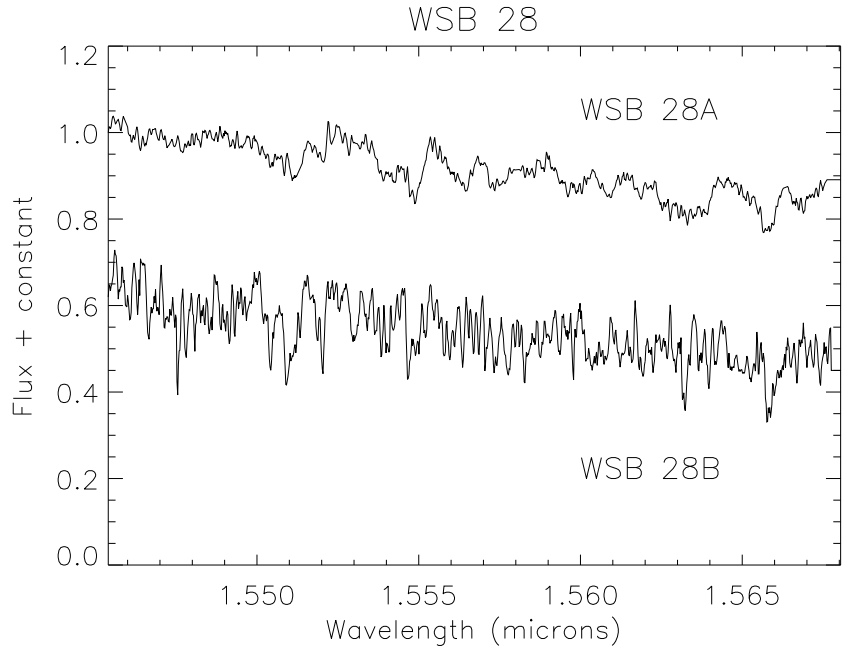}
%\picplace{5cm}{2cm} % Give the correct figure height and width in cm
%
\caption{R$=$30,000 spectra of the components in the young binary
WSB 28.  The $v$sin$i$'s are discrepant by a factor of 2$-$3,
indicating either unaligned rotation axes or significantly
different  rotation periods.  Veiling from circumstellar material
cannot account for the shallow features in the M3 primary
because this component is not associated with any circumstellar
material, although the M7 secondary is \cite{mcc06}).}
\label{fig:1}       % Give a unique label
\end{figure}

How much of an impact might selection effects have on the
results presented here?
Certainly small mass ratio systems are more difficult
to detect as well as to characterize, particularly in the
most interesting small separation regime (\S1).
Systems classified as weak-lined T~Tauris, unresolved, might also 
harbor truncated disks around the secondary stars.  Such small
structures could go undetected as the result of dilution from a
relatively bright primary.  Circumstellar disks with central
holes that show excesses in the mid-infrared but not in the 
near-infrared, and which do not show signatures of accretion,
may be present but are effectively undetectable.  Even if 
sensitive but low-angular resolution \emph{Spitzer} observations
could reveal the presence of such a structure, there is little
recourse for ground-based mid-infrared follow up at
sufficiently high sensitivity and angular resolution.
Only 4 of the circumstellar disks in the young binary sample
of \cite{mcc06} (TTau N and S, UZ Tau E, and RW Aur A)
are brighter than the N$=$4~mag limit of the VLTI mid-infrared
instrument MIDI.

We must also take into account that
the completeness of our knowledge of binary populations
varies markedly between different star forming regions, possibly
leading to an inaccurate determination of differences in mixed
pair fractions, etc, between regions.  Taurus, given its small
size and ready accessibility in the northern skies, is arguably
the most thoroughly studied region.  However, its
faintest members are only now being surveyed for multiplicity
\cite{kon07}.

\section{Outer Disks}
\label{sec:3}

The cool gas and dust in outer disks, including circumbinary disks, is
best surveyed using far-infrared or submillimeter observations. Disks
are usually optically thin at long wavelengths, so these observations
have the additional benefit of providing total disk masses
(e.g. \cite{AndreMontmerle94}) in the region analogous to where giant
planets formed in the Solar System.

Although estimates of the binary fraction were highly incomplete when
the first submillimeter surveys were done, it was still clear
immediately that binary stars with separations closer than 100 AU were
deficient in disks \cite{Jensen1994,OsterlohBeckwith95}.  In recent
work, a survey of 150 young stars in Taurus including 62 multiple stars
showed lower submillimeter fluxes and hence masses in systems closer
than $\sim$100 AU than in single stars, while wide binaries were similar
to single stars in disk mass \cite{AndrewsWilliams2005}. Disks were
present, albeit at these lower masses, in approximately the same
fraction of multiple star as single star systems. Perhaps these disks
can still form giant planets, but of lower average mass, than the single
stars.

Models of disk dissipation generally show that the circumsecondary disk, which should be truncated closer to the star due to the primary, should dissipate faster \cite{Armitage1999}.  In single stars, disk mass is not correlated with stellar mass \cite{AndrewsWilliams2005}, so it is quite possible for circumsecondary disks to start out as more massive than circumprimary disks, and these initial conditions can overwhelm the difference in dissipation timescale.

The surveys described above were carried out with single dish telescopes
and therefore have low spatial resolution incapable of distinguishing
primary and secondary disks at the interesting separation range of
$<\sim$100 AU.  A smaller number of objects have been surveyed with
interferometers that can resolve the multiple systems. In one such
survey, the primary stars of four binaries in Ophiuchus hosted higher
mass disks, even when the secondaries were still accreting, while in four binaries
in Taurus the circumsecondary disks were more massive
\cite{Patience05}. In these very young objects, the true ``primary'' may
have been misidentified in extincted visual-wavelength data, or these
trends may relate to the initial conditions. For four wider systems,
also in Taurus, the circumprimary disks were again the most massive (and
again comparable to single stars in Taurus) \cite{JensenAkeson03}.

\section{Orientation of Disks in Young Binaries}
\label{sec:4}

A single star plus disk system contains a
single plane:  that of the disk.  A binary system, however,
is associated with four relevant planes:  a circumstellar disk around
each star, the plane of the binary orbit, and the plane of
any circumbinary disk, although the latter are relatively rare
\cite{jen97}.  \emph{Alignment} of circumstellar disks does not
necessarily imply coplanarity of the binary orbital plane with that
of the aligned disks (Figure 2).  The polarization studies
of \cite{jen04} and \cite{mon06} trace circumstellar alignment,
for relatively wide, angularly resolved young binaries,
using the orientation of the polarization position angles
of dust grains in the disks.  They found that most simple
binary systems studied exhibit aligned disks with polarization position
angles consistent to within $<$30 degres, although higher order
multiples show a large range of variation in polarization position
angles. 

\begin{figure}
\centering
\includegraphics[height=9cm]{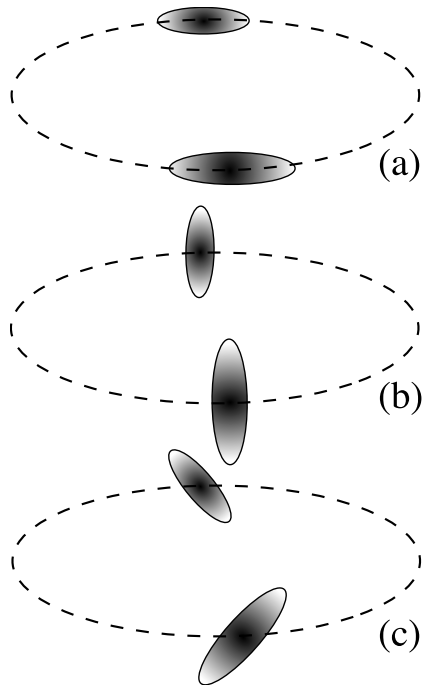}
%\picplace{5cm}{2cm} % Give the correct figure height and width in cm
%
\caption{Circumstellar disks in a simple binary orbit:
(a) aligned and coplanar, (b) aligned but non-coplanar,
(c) unaligned and non-coplanar.}
\label{fig:2}       % Give a unique label
\end{figure}

The orientations of the highly collimated jets that
emanate from many young star systems provide a proxy for
determining disk orientations in unresolved binaries as jets
are thought to launch perpendicular to the inner
circumstellar disks.  Multiple misaligned
jets are known to exist in a number of young systems (e.g.,
\cite{mon07} and refernces therein), suggesting that it is
possible for small separation binaries to actually \emph{form}
with misaligned disks (Figure 2, case c).  Thus,
formation models must account for this counterintuitive evidence.

The coplanarity of disks and binary orbits is readily studied for
some well-separated pairs.  Interestingly, it appears likely that
circumbinary disks are aligned with close binary star orbits,
e.g., for DQ Tau, UZ Tau E, and HD 98800 B \cite{mat97, pra02,
pra01}.  However,
systems with a circumstellar disk around at least one component
of a wider binary, e.g., HV Tau AB-C, HK Tau A-B, UZ Tau E-W,
T Tau N-S, and HD 98800 N-S \cite{sta03, sta98, pra02,
ake02, pra01}, \emph{do not} appear to 
be coplanar.  The dynamics of circumbinary and even close
circumstellar disks and the interrelationship between disks
and orbits appears to be complex and is not yet well
understood.  We present these conclusions as a cautionary
tale:  even binaries with separations of a few tens of
AU -- or less -- cannot be assumed to harbor aligned disks
coplanar with binary orbits.  In higher order multiples,
misalignments may be the rule.  It is likely that in at least
some cases misalignment may have its origins in the formation
dynamics of these systems.

\section{Debris Disks and Binaries}
\label{sec:5}

There are several well studied examples of binary systems amongst the
older class of circumstellar disks -- the transitional and debris
systems. These are disks in which the primordial material, particularly
gas, is partially or totally dissipated and remaining solids are large
enough that their major destruction mechanism is collisions (either
aggregation onto planets or disruptive).  Giant planets must either have
already formed in these systems or will not and terrestrial planets may
be in their final stages of accumulation, perhaps eras akin to the late
heavy bombardment in the solar system.  These systems are closer to the
Sun than the nearest sites of recent/ongoing star formation discussed
earlier, so the affect of binarity on the disks can be observed in some
detail. We will discuss two examples.

HD 141569 is a hierarchical triple star system consisting of an A0-type
primary star, which sports an extended disk containing both small
quantities of gas and dust, and two M-type companion stars located about
1000 AU away. The low mass stars, and presumably
the whole system, are about 5 Myr old \cite{Weinberger00}.
Spiral structure at 200--500 AU in
the primary's disk can be explained by either a highly eccentric (e
$>\sim$ 0.7) binary A-BC orbit \cite{AugereauPapaloizou04,Quillen05,Ardila05} or a recent
($\sim$ 1000 yr ago) stellar flyby \cite{Beust05}.  In both cases, the
portion of the disk affected by the companions is at distances of a few
hundred AU,and structure in the disk at $<$150 AU must have another
cause, perhaps a planet.  Interestingly, the two M-type stars have no
detectable disks below the level seen around the primary star. This
could be due to the small separation of their orbit, $\sim$150 AU.

HD 98800 is a member of the $\sim$8 Myr old TW Hya Association and is
composed of four nearly identical stars in two spectroscopic
binaries. All of the dust encircles one of these pairs, HD 98800Bb
\cite{Low99,koe00,pra01}.Thus, the system has characteristics of both a
circumbinary and circumstellar disk.  The Bb binary is eccentric
(e=0.78) with a semi-major axis of 1 AU \cite{bod05}. Based on its
temperature, the inner edge of the dust disk sits at 1.2 -- 2.1 AU
(Prato et al.). This
is just barely consistent with estimates of the dynamical tidal
truncation \cite{ArtymowiczLubow94}. The A-B orbit is also significantly
eccentric (0.3--0.6) with a periastron approach of perhaps 35 AU
\cite{Tokovinin99}. The outer edge of the dust disk is less well
constrained by the infrared/submillimeter observations, but is $>$5 AU
and could be as large as 25 AU \cite{koe00}.  An outer size of 10 AU would
fit both the observations of the dust temperature and the expected
dynamical truncation due to the A-B orbit.

While both of these systems provide interesting examples of the
dynamical influence of multiplicity on the disk, they also illustrate
that planet formation is possible under such complicated
circumstances. The small dust grains in the HD 141569A and HD 98800B
disks are regenerated in collisions
\cite{Weinberger99,AugereauPapaloizou04,Low99} and indicate that
planetesimals did form on timescales short enough that gas could have
been present simultaneiously with solid bodies..

Statistics of the incidence of debris disks around binaries are
consistent with the idea that wide binaries do not affect disk
evolution. A survey of 69 FGK stars including binaries of separations
$>$ 500 AU finds 3/8 of the disks are around binary members \cite{Bryden06}.

\section{Future Tests and Observations}
\label{sec:6}

A tremendous observational effort is required to explore
the most populous binary separation regime, and that of most scientific
interest with respect to planet formation --- a few to $\sim$30~AU
separations.  

Ongoing spatially resolved spectroscopy with adaptive optics systems on
large telescopes will assess the accretion parameters and inner disk
optical depths of circumstellar disks in close binaries. With concerted
observational attention, it seems a solvable problem to measure the
dissipation timescales of primary and secondary disks. Ground based
interferometers will get detailed orbits for close binaries which can
then be compared to disk sizes for empirical verification of dynamical
estimates of tidal disruption and dissipation.

One big advance will come with operation of the Atacama Large Millimeter
Array (ALMA). With its sub-arcsecond resolution, comparable to that of
Hubble Space Telescope, it will be able to determine the masses and
orientations of the circumstellar disks in binary systems at the
critical separation range.

%\begin{table}
%\centering
%\begin{tabular}{lll}
%\hline\noalign{\smallskip}
%\end{tabular}
%\end{table}

%
%
%
% BibTeX users please use
% \bibliographystyle{}
% \bibliography{}
%
% Non-BibTeX users please follow the syntax
% the syntax of "referenc.tex" for your own citations
i%%%%%%%%%%%%%%%%%%%%%%%% referenc.tex %%%%%%%%%%%%%%%%%%%%%%%%%%%%%%
% sample references
% "physics"
%
% Use this file as a template for your own input.
%
%%%%%%%%%%%%%%%%%%%%%%%% Springer-Verlag %%%%%%%%%%%%%%%%%%%%%%%%%%

%
% BibTeX users please use
% \bibliographystyle{}
% \bibliography{}

%%%%%%%%%%%%%%%%%%%%%%%%%%%%%%%%%%%%%%%%%%%%%%%%%%%%%%%%%%%%%%%%%%%%%%  }

%%%%%%%%%%%%%%%%%%%%%%%%%%%%%%%%%%%%%%%%%%%%%%%%%%%%%%%%%%%%%%%%%%%%%%

\printindex
\end{document}